\documentclass[man,floatsintext]{apa7}

\usepackage[american]{babel}

\usepackage{csquotes}

\usepackage{amsmath, amssymb}
\usepackage{booktabs}
\usepackage{threeparttable}
\usepackage{makecell}
\usepackage{multirow}

\usepackage{enumitem}

\usepackage{xcolor}
\usepackage{ulem}

\usepackage[style=apa,sortcites=true,sorting=nyt,backend=biber]{biblatex}

\newcommand{\IQR}{\mathit{IQR}}

\DeclareLanguageMapping{american}{american-apa}
\title{Dynamical incompatibilities in paced finger-tapping experiments}
\shorttitle{Dynamical incompatibilities in paced finger-tapping experiments}

\authorsnames[{1,2},1,{1,3,4}]{Ariel D. Silva, Claudia R. Gonz\'alez, Rodrigo Laje}
\authorsaffiliations{{Universidad Nacional de Quilmes, Departamento de Ciencia y Tecnolog\'{\i}a, Sensorimotor Dynamics Lab, Argentina},{CONICET, Argentina},{Universidad de Buenos Aires, FCEN, Departamento de Computaci\'on, Argentina},{CONICET-Universidad de Buenos Aires, Instituto de C\'alculo (IC), Argentina}}

\leftheader{Silva, Gonz\'alez and Laje}

\abstract{Paced finger-tapping tasks are used to probe the error correction mechanism underlying sensorimotor synchronization. Despite their century-long history, fundamental contradictions persist in the literature. One such contradiction arises when comparing the two most common types of period perturbation: step change and phase shift. The stimulus sequence is exactly the same up to and including the (unexpected) perturbed stimulus. Why then would the timing of the next response be different between perturbation types, as observed? We show, both experimentally and theoretically, that responses to both types of perturbation are dynamically incompatible when recorded in separate experiments; that is, they cannot be described by a single underlying dynamical system due to the build-up of different temporal contexts. In contrast, when both types of perturbation are presented randomly within the same experiment, the responses become compatible and can be explained by a single mechanism. We conclude that a single underlying dynamical system can represent the response to all perturbation types, signs, and sizes, which is nevertheless calibrated by temporal context. Our results challenge the established idea of phase and period correction processes that are separately activated for different perturbation types.

Word count: Approximately 6800

{\bf Public Significance Statement}

Temporal processing is essential for many human behaviors, including coordination with external rhythms in activities such as music and dance. Finger-tapping tasks are widely used to study these processes and have also been employed to investigate motor control disturbances in neurological disorders such as Parkinson’s disease. This study analizes how people resynchronize after unexpected tempo perturbations. The evolution of the temporal error correction during resynchronization depends on the temporal context in which these perturbations occur, and it is shown that responses to different types of perturbations in pure context are incompatible with each other. These findings help explain why previous studies have produced seemingly contradictory results and provide a more accurate basis for interpreting and modeling human sensorimotor synchronization.
}

\authornote{This work was supported by Universidad Nacional de Quilmes (Argentina, grant \#53/1028) and CO\-NI\-CET (Argentina).

The authors declare no competing interests.

Correspondence concerning this article should be addressed to Rodrigo Laje, Departamento de Ciencia y Tecnolog\'{\i}a, Universidad Nacional de Quilmes, R.\ S.\ Pe\~na 352, Bernal B1876BXD, Argentina. E-mail: rlaje@unq.edu.ar
}

\keywords{Sensorimotor synchronization; Period perturbations; Dynamical systems; Discrete time; Difference embedding}

\begin{document}
\maketitle

In the study of time processing and temporal cognition at the scale of hundreds of milliseconds, the last two decades have brought significant advances in describing behavior and understanding their neural bases and developmental stages \parencite{Tsao2022,Hogendoorn2022,Monteiro2023,Paton2018,Buzi2024}. Recognizing that time is processed on multiple time scales is key to understanding how the brain and body coordinate during movement planning and control \parencite{Inagaki2022}. Within this time scale lies sensorimotor synchronization (SMS), that is, the ability to adjust our movements to match a periodic external stimulus that is the basis of human activities such as music and dance \parencite{Repp2005,Repp2013}. A limited form of SMS has also been recently demonstrated in non-human animals like monkeys \parencite{Rajendran2025} and rats \parencite{Ito2022}. Research on SMS as a potential diagnostic marker in neurodegenerative diseases has seen a rise in the last decade \parencite{Roalf2018,Schnehen2022} for diseases where motor dysfunction is a primary symptom like Parkinson's \parencite{Surangsrirat2022,Williams2023} and even others where motor deficits occur later, like Alzheimer's \parencite{Koppelmans2022,Koppelmans2023} and mild cognitive impairment \parencite{Ilardi2022,Rudd2023}.

The most basic paradigm for investigating SMS is paced finger tapping, in which subjects are asked to synchronize to an external metronome (usually a periodic sequence of brief tones). Within this framework, researchers often introduce unexpected changes to the interstimulus interval (ISI) to study how the system responds. This task has enabled key advances, such as the identification of perceptual and motor components involved in the resynchronization process following a perturbation \parencite{Rosso2023}, as well as the observation of an asymmetry in the way lead and lag synchronization errors are processed in the brain, consistent with a previously documented behavioral asymmetry \parencite{Bavassi2017}. The experimental paradigm of period perturbations in SMS is used not only to study human synchronization but also synchronization of non-human animals \parencite{Rouse2016} and mechanisms of broader cognitive significance like reading fluency \parencite{Wei2022}.

It is generally accepted that average synchronization and its recovery after a perturbation are produced by an error correction mechanism \parencite{Repp2005}. It has been further proposed that the error correction mechanism is composed of two distinct processes: phase and period correction \parencite{Harry2023,Repp2004,Repp2008,Repp2013,vanderSteen2013}. These two putative processes are thought to have different time scales and to depend differently on attention. Importantly, it is assumed that they are separately activated depending on the perturbation type presented to the participant, an assumption that is specifically reflected in the way models are fitted to experimental data---the equation representing period correction is not included in the fitting when phase-shift-like perturbations are presented \parencite{Harry2023}. How this theoretical framework accounts for the selective activation of the two processes depending on the type of the upcoming (and unexpected) perturbation remains unresolved.

Although linear models have long been used to describe this resynchronization process \parencite{Repp2013}, more recent analyses of time series under different perturbation types and sizes suggest that this mechanism exhibits nonlinear characteristics \parencite{Bavassi2013,Bavassi2017,Lopez2019,Gonzalez2019,Silva2024}. Other experimental results also support the need to abandon linear models in favor of more complex descriptions \parencite{Haken1985,Loehr2011,Egger2020,Large2023,Lainscsek2012,Roman2019,Konvalinka2009,Heggli2019,Slowinski2024}. This change of perspective could have relevant implications for experimental design, the methodologies used for data analysis, and the development of diagnostic tools \parencite{Lopez2019,Silva2024}.

Despite this progress, key questions remain unresolved. In this work, we focus on a longstanding contradiction in the literature of finger-tapping experiments. That is, the participant's response following an unexpected period perturbation is different for different perturbation types, despite the fact that the stimulus sequence is the same up to and including the first perturbed stimulus \parencite{Silva2024}. To explain this contradictory observation, we reanalyze SMS data \parencite{Silva2024} and show that responses to different perturbation types are experimentally incompatible with each other in terms of a putative underlying dynamical system, unless they are presented in the same experiment. We also use a reduced version of a previously developed nonlinear model \parencite{Gonzalez2019} to fit the data and theoretically demonstrate this incompatibility. Our results allow us to reconcile contradictory experimental results in the literature and reinterpret the historical difficulty that mathematical models of SMS have had in reproducing, with a single set of parameters, the response to perturbations from different experiments. Our work warrants a revision of the theoretical account of phase and period correction processes that are selectively activated for different perturbation types.

\section{Methods\label{sec.methods}}

\subsection{Original dataset}

The experimental data we reanalyze here were previously published \parencite{Silva2024}. For the sake of completeness we reproduce the most important parts of the original methods.

\subsubsection{Experimental data and task description}

The objective for the original dataset \parencite{Silva2024} was to determine whether the perturbation context influences the resynchronization phase following a period perturbation in a paced finger-tapping task. Participants ($N=71$; 33 women; all right-handed; ages 18-63) were instructed to tap with their fingers to an auditory rhythm and maintain average synchrony. Sample size of the original dataset was determined by a priori power analysis \parencite{Silva2024}. See below for a discussion of sample size vs qualitative feature comparisons.

A trial consisted of a sequence of 35 brief tones with a baseline interstimulus interval $T=500$ ms and their responses. Occurrence times of responses $R_n$ and stimuli $S_n$ were recorded. The basic experimental data was the observed asynchrony $e_n$, that is the time difference between corresponding occurrences of responses $R_n$ and stimuli $S_n$:
\begin{equation}
e_n = R_n - S_n
\label{eq.asyn}
\end{equation}

\noindent where $n$ is the response/stimulus number along the sequence.

Upon the appearance of a period perturbation (i.e., an unexpected change to the baseline interstimulus interval), participants were instructed to regain synchrony without interrupting the tapping. A period perturbation could occur at a randomly chosen step $n$ in the range 17-22. Steps were renumbered afterwards such that the perturbation occurs at $n=0$.

Two perturbation types were used \parencite{Repp2005}: step changes (SC), where the stimulus period is abruptly modified by an amount $\Delta T$ at a random step of the sequence; and phase shifts (PS), where the period changes at two consecutive steps, first by $+\Delta T$ and then by $-\Delta T$, thus returning to its original value. Perturbations could be positive ($\Delta T > 0$, “pos”) or negative ($\Delta T < 0$, “neg”), with a magnitude of either 20 or 50 milliseconds.

Critically, the subjects were grouped into two contexts. In the “pure” context participants were exposed to only one type of perturbation (Group 1, SC perturbation; Group 2, PS perturbation; both groups with 20 and 50 ms perturbation magnitudes). In the “combined” context participants were exposed to both types of perturbation, randomly interspersed on a trial-by-trial basis (Group 3, 50 ms magnitude; Group 4, 20 ms magnitude).

The experimental design was fully factorial, with factors Context (pure and combined) x Type (SC and PS) x Sign (pos and neg), with additional isochronous control conditions ($\Delta T = 0$).

\subsubsection{Data exclusion}

Outliers were treated exactly as described in our previous work where the original data were recorded \parencite{Silva2024}, as described in the following. No further additional exclusion criteria were applied.

We used a robust definition of outliers (Tukey’s fences) and applied uniform criteria at the trial, participant, and condition levels:
\begin{itemize}[topsep=0pt, partopsep=0pt, itemsep=0pt, parsep=0pt]
\setlength{\itemsep}{0pt}
\item Trial level
    \begin{itemize}[topsep=0pt, partopsep=0pt, itemsep=0pt, parsep=0pt]
    \setlength{\itemsep}{0pt}
    \item Trial mean asynchrony. For every participant and condition, we computed the mean asynchrony of every trial, and found the lower ($Q_1$) and upper ($Q_3$) quartiles of the distribution. Any trial with a mean asynchrony outside the interval $[Q_1 - 1.5 \IQR; Q_3 + 1.5 \IQR]$ (Tukey’s fences, where $\IQR = Q_3 - Q_1$ is the interquartile range) was flagged as outlier and removed from the dataset.
    \item Trial standard deviation of asynchronies. For every participant and condition, we computed the standard deviation of the asynchronies of every trial, and lower and upper quartiles of the distribution. Any trial with a standard deviation outside Tukey’s fences was flagged as outlier and removed from the dataset.
    \end{itemize}
\item Participant level
    \begin{itemize}[topsep=0pt, partopsep=0pt, itemsep=0pt, parsep=0pt]
    \setlength{\itemsep}{0pt}
    \item Participant in a condition. If more than 50\% of trials from the same participant were removed as outliers in a given condition, then the whole participant was removed from the condition.
    \item Participant mean asynchrony. We computed the mean asynchrony of every participant in a condition, and the lower and upper quartiles of the distribution. Any participant with a mean outside Tukey’s fences was flagged as outlier and removed from the condition.
    \item Participant standard deviation of asynchronies. We computed the standard deviation of asynchronies of every participant in a condition. Any participant with a standard deviation outside Tukey’s fences was flagged as outlier and removed from the condition.
    \end{itemize}
\item Condition level
    \begin{itemize}[topsep=0pt, partopsep=0pt, itemsep=0pt, parsep=0pt]
    \setlength{\itemsep}{0pt}
    \item Condition in the experiment. If a participant was removed as an outlier in more than 50\% of the conditions, then he/she was removed from the whole experiment. One participant was removed from Group 1, and two were removed from Group 2.
    \end{itemize}
\end{itemize}
The highest number of outlier trials per participant was 11, representing 15\% of his/her total trials (at least 1 outlier trial was detected for every participant). The percentage of outlier trials in Groups 1 through 4 was 8.5\%, 9.3\%, 8.1\%, and 7.9\%, respectively.

\subsubsection{Ethics approval}

The experimental protocols were designed in accordance with national and international guidelines and were approved by the Ethics Committee of the Universidad Nacional de Quilmes (approval \#2013-06-14). Written informed consent to participate and to publish non-sensitive anonymized data was obtained from all participants.

\subsection{\label{sec.embed_methods}Reconstruction of the experimental phase space via embeddings}

In this work we show the dynamical incompatibility between responses by analyzing trajectory crossings in phase space using the embedding technique. Previous work showed that the error-correction mechanism underlying resynchronization after a period perturbation is best described by a dynamical model with two variables \parencite{Bavassi2013,Gonzalez2019}. The embedding technique \parencite{Gilmore1998,Lopez2019} allows us to reconstruct a 2D phase space from the experimental time series of the asynchrony $e_n$.

The variable $e_n$, however, is ill-defined when the interstimulus interval changes \parencite{Gonzalez2019}. Based on our previous work, we switch to a variable that is more appropriate in the presence of period perturbations: the predicted asynchrony $p_n$, defined as
\begin{equation}
p_n = e_n + \Delta T
\label{eq.predicted}
\end{equation}

\noindent where $\Delta T$ is the size of the period perturbation. This simple relationship between $e_n$ and $p_n$ is best conceptualized by realizing that, when an unexpected period perturbation $\Delta T$ occurs at step $n$ in the sequence, the asynchrony value predicted by the participant $p_n$ will not be the same as the actually observed value $e_n$ due to the experimental manipulation \parencite{Gonzalez2019}.

After switching to the more appropriate variable $p_n$ as described above, we defined the following difference embedding: predicted asynchrony ($p_n$) on the horizontal axis vs difference between consecutive predicted asynchronies ($p_n - p_{n-1}$) on the vertical axis, averaged across subjects. The post-perturbation baseline was subtracted from all values to show convergence to the new equilibrium.

In order to prevent visual cluttering, we avoid plotting pre-perturbation and post-perturbation baseline data and instead plot data from the resynchronization phase only (SC: $1 \leq n \leq 8$; PS: $2 \leq n \leq 8$).

\subsection{\label{sec.model}Mathematical model}

We model the experimental data with a single nonlinear dynamical system proposed in a previous work \parencite{Gonzalez2019}. In this work we simplified it to include two nonlinear terms only:
\begin{equation}
\begin{aligned}
p_{n+1} & = a e_n + b (x_n - T_n) + \beta e_n (x_n - T_n)^2 \\
x_{n+1} & = c e_n + d (x_n - T_n) + \delta e_n^2 + T_n
\end{aligned}
\label{eq.model_methods}
\end{equation}

\noindent where $e_n$ is the actually observed asynchrony at step $n$ (Eq. \ref{eq.asyn}); $p_n$ is the predicted asynchrony (Eq. \ref{eq.predicted}); and $x_n$ is a variable of dynamical origin needed to reproduce the behavior \parencite{Bavassi2013,Gonzalez2019}. The metronome is represented by the interstimulus interval $T_n$ whose value can be set by the experimenter. An SC perturbation occurring at $n=0$ changes the metronome period once at $n=0$ ($T_n=T_{base}$ for $n<0$ and $T_n=T_{base} + \Delta T$ for $n \geq 0$); a PS perturbation changes it twice ($T_n=T_{base}$ for $n<0$, $T_n=T_{base} + \Delta T$ at $n=0$, and back to $T_n=T_{base}$ for $n \geq 1$) \parencite{Bavassi2013}.

Our choice of a nonlinear model is based on previous works showing clearly nonlinear aspects in the experimental data \parencite{Bavassi2013,Bavassi2017,Lopez2019,Gonzalez2019,Silva2024} and on the following rationale. The way researchers in the finger tapping literature traditionally overcome the shortcomings of linear models is to fit the model separately for every condition (e.g., every perturbation size, sign, and type), leading to models that have an effective number of parameters much larger than originally presented and that are thus much less parsimonious (see our review in Ref.\ \parencite{Gonzalez2019}. It would be very difficult to determine compatibility/incompatibility based on such a fitting procedure, where fitted parameter values can change not only between contexts but also between perturbation signs and sizes. It would probably lead to all incompatible situations, with theoretical and experimental conclusions diverging from each other.

\subsection{\label{sec.fitting}Model fitting}

The model was fitted to the data using the Differential Evolution function from the SciPy library in Python. We defined the fitting loss function as the Euclidean distance between the experimental time series and the corresponding model time series:
\begin{equation*}
\mathit{dist} = \sqrt{\sum_{n=-5}^{11} \left(e_n^{(exp)} - e_n^{(model)} \right)^2}
\end{equation*}

\noindent where $e_n^{(exp)}$ represents the experimental values of the asynchrony and $e_n^{(model)}$ the values predicted by the model at every step $n$ in the sequence.

All six model parameters were set free for fitting within the following bounds: $a, b, c, d \in [-2, 2]$; $\delta \in [-0.02, 0.02]$; $\beta \in [-0.0002, 0.0002]$. The order of magnitude of the bounds (linear $\sim1$, quadratic $\sim0.01$, cubic $\sim0.0001$) was determined by dimensional considerations. We checked afterwards that the obtained distributions were not artifactually truncated by the set bounds.

Hyperparameter values for Differential Evolution were: $\mbox{mutation} = (0.5, 1)$, $\mbox{recombination} = 0.7$, $\mbox{polish} = \mbox{False}$, $\mbox{workers} = -1$, $\mbox{maxiter} = 200$, $\mbox{popsize} = 100$, $\mbox{tol} = 0.03$, $\mbox{disp} = \mbox{True}$.

Our model doesn’t include a constant baseline that is usually found in the experimental data \parencite{Repp2005,Bavassi2013}. In order to improve fitting, a constant baseline was added to the model variable $p_n$ in each condition. Its value is equal to the average asynchrony in the preperturbation region ($n < 0$) and the average asynchrony in the postperturbation region ($n \geq 7$). For plotting purposes, the preperturbation baseline was shifted to zero by subtracting its average value.

Model fitting was performed as follows, according to our rationale in the Results section:
\begin{table}[!ph]
\begin{tabular}{ll}
\parbox[t]{0.6\columnwidth}{\bf Perturbation type and context}
& \bf Observations \vspace{0.2cm} \\ \hline
SC pure, PS pure & Separate fittings \\
SC comb, PS comb & Separate fittings \\
SC pure, PS pure & Joint fitting \\
SC comb, PS comb & Joint fitting \\
\end{tabular}
\end{table}

In each case, the two perturbation signs (pos/neg) and the two perturbation sizes (20 ms/50 ms) were included in the fitting.

\subsection{\label{sec.dists}Parameter distributions}

Two hundred independent fittings were performed for each case (see previous subsection), and the distributions of the resulting parameters were analyzed. In particular, the distributions of parameter $a$ were studied for the conditions SC in pure context, PS in pure context, joint fitting for PS and SC in pure context, SC in combined context, PS in combined context, and joint fitting for PS and SC in combined context.

For the SC in pure context, joint fitting for PS and SC in pure context, and joint fitting for PS and SC in combined context cases, the fitted values of parameter $a$ displayed a bimodal distribution that we associate with two distinct subpopulations (not shown). The boundaries between the two subpopulations were established at $a=0.4$, $a=0.75$, and $a=0.8$, respectively.

We removed outliers of parameter $a$ distributions by using the standard Tukey’s fences algorithm: a value is detected as outlier if it is outside the interval $[Q_1 - 1.5 \IQR; Q_3 + 1.5 \IQR]$, where $Q_{1,3}$ are the first and third quartiles of the distribution and $\IQR$ is the interquartile range. The same criteria were used to detect outliers in the $\beta$ parameter for PS in pure context and SC in pure context.

\subsection{\label{sec.overlap}Distribution overlap}

The resulting subpopulations were plotted together (SC in pure context vs PS in pure context and SC in combined context vs PS in combined context). We defined two perturbation types as compatible if their parameter distributions after separate fittings are overlapped to some extent. To quantify the overlap, the $O_{obs}$ index was defined, which measures the percentage of common values between two populations of the same parameter, belonging to different types of perturbation within the same context.

The $O_{obs}$ calculation considered the common range of values between both distributions, divided into 30 bins. The total number of data points with common bins ($total\_common$) was counted and normalized by the total number of data points ($total\_data$): $O_{obs} = (total\_common / total\_data) \times 100$.

\subsection{\label{sec.distance}Distribution of distances}

The distributions of the variable $\mathit{dist}$ were plotted, especially for the joint fitting for PS and SC in pure context and joint fitting for PS and SC in combined context conditions (see Figure \ref{fig.distance}). To evaluate the significance of the differences between the two populations, a two-tailed Mann–Whitney–Wilcoxon test was applied. The following were defined: Pure group ($\mathit{dist}$ values from the pure context); Combined group ($\mathit{dist}$ values from the combined context); $n_1$ and $n_2$, respective sample sizes. The Mann-Whitney $U$ statistics and $p$-value were calculated, as well as the appropriate effect size $\delta_\mathit{Cliff} = 2 U / (n_1 n_2) - 1$. A bootstrap was performed with 10,000 iterations to estimate the mean of $\delta_\mathit{Cliff}$ and its 95\% confidence interval (CI).

\subsection{Sample size vs qualitative feature comparisons}

The number of participants in the original dataset was determined by a priori power analysis as described in our previous work \parencite{Silva2024}. The hypotheses we test here, however, are qualitative in nature. Our main result is based on a qualitative comparison between the shapes of different trajectories in phase space: whether they show clear crossings or they can be construed as smoothly merging. This kind of reasoning about the geometrical organization of trajectories in phase space is a well-established tool in Dynamical Systems theory \parencite{Gilmore1998}. This qualitative feature comparison can be described as exploratory and doesn't have an associated effect size. The analyses were intended to characterize dynamic incompatibility and parameter space structure and thus were exploratory and qualitative in nature. The main conclusions are based on the consistency of data- and model-based dynamical patterns, rather than null hypothesis significance testing.


Regarding the additional, quantitative results that support our conclusion (model fittings, parameter distributions, etc), we performed 200 independent repetitions, a number high enough to have well-defined distributions with very few outliers, and still practical from the point of view of calculation time. Again, due to the exploratory nature of our work, no effect size of interest was defined.

\subsection{Transparency and openness}

All data exclusions, manipulations and analyses performed in this study are reported either here or in a previous publication \parencite{Silva2024}. The data and code to reproduce all analyses and figures are publicly available on OSF (\url{https://osf.io/bazhp}) and GitHub (\url{https://github.com/SMDynamicsLab/Incompatibilities2026}). The analyses were performed in Python. The design and analysis of this study were not pre-registered.

\subsection{Colorblind-friendly and grayscale-compatible plots}

We used the plasma colormap.

\section{\label{sec.results}Results}

Our main research question is whether the responses to SC and PS perturbations in pure contexts are compatible with each other; that is, whether they can be produced by the same underlying dynamical system. In the following we present converging evidence from experimental and theoretical results showing that the responses are incompatible.

\subsection{\label{sec.embeddings}Experimental time series from pure contexts are incompatible with each other}

\begin{figure}[!pt]
\includegraphics[width=\textwidth]{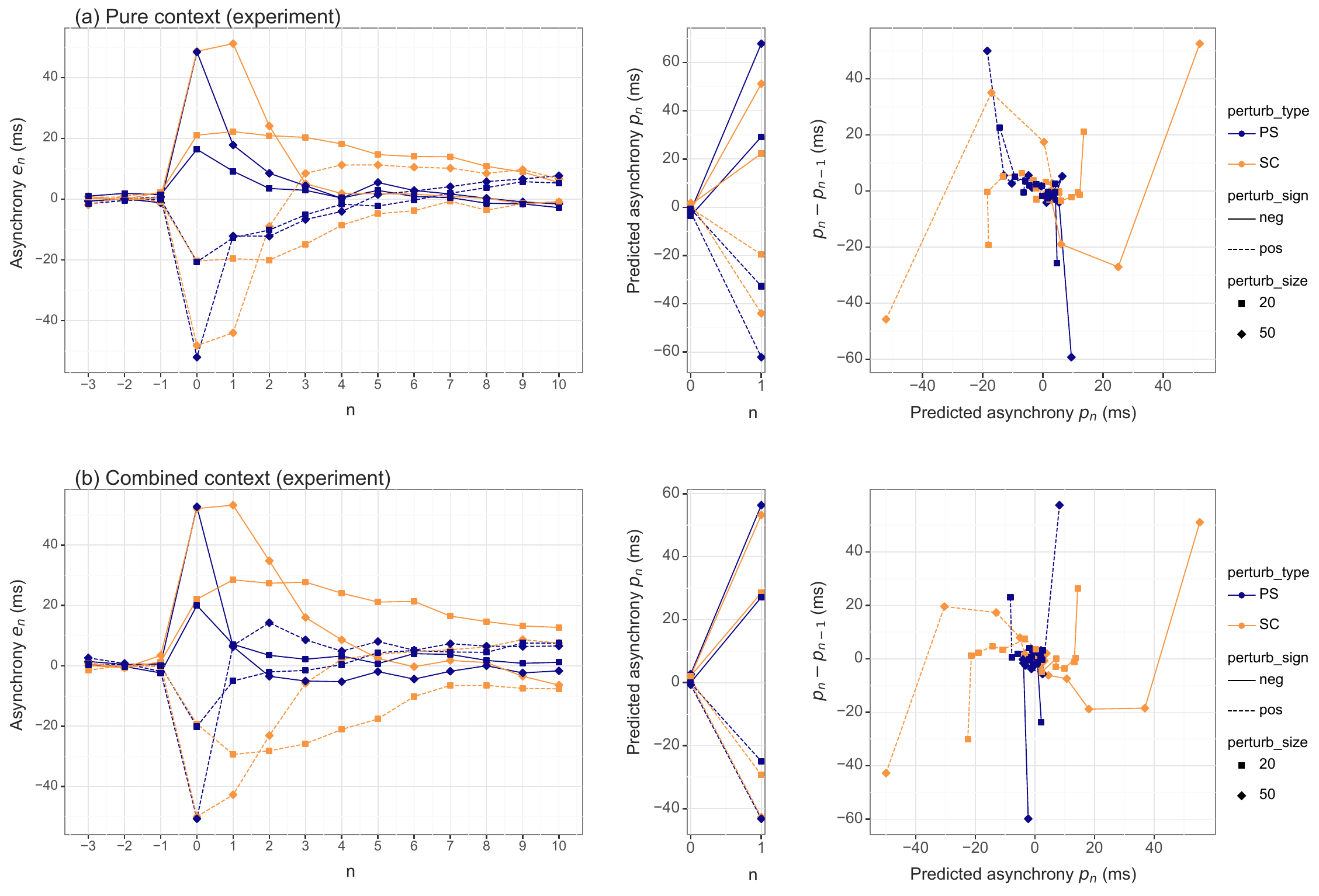}
\caption{Crossings in the reconstructed phase space point to an incompatibility between SC and PS from pure contexts. (a) Data from pure context. Left: experimental time series. The (unexpected) perturbations occur at $n=0$; the forced error is reflected in a large asynchrony value, on average equal to the opposite of the perturbation size $\Delta T$. Participants gradually recover synchronization afterwards. Middle: Predicted asynchrony after perturbation. SC (orange) and PS (blue) curves have different futures ($n=1$) despite coming from similar values at $n=0$. Right: Difference embeddings of the time series (SC: $1 \leq n \leq 8$; PS: $2 \leq n \leq 8$). Trajectories converge to the origin during resynchronization. A clear crossing of trajectories between SC (orange) and PS (blue) occurs in the top left quadrant of the phase space, pointing to a dynamical incompatibility between them. (b) Data from combined context. The predicted asynchrony (middle panel) doesn't show divergence between blue and orange curves and  trajectories (right panel) merge smoothly with no clear crossings above noise level. Mean across participants; error bars not shown for clarity.
\label{fig.embeddings}}
\end{figure}

Figure \ref{fig.embeddings} shows how participants resynchronize after SC and PS perturbations in pure context (top row) vs combined context (bottom row). In the left panels we plot the asynchrony $e_n$ as a function of step $n$ in the sequence (Eq.\ \ref{eq.asyn}). The unexpected perturbations occur at $n=0$ where the asynchrony takes a large value due to the forced error; participants return to baseline after resynchronization.

The dynamics of these time series are, however, incompatible with each other. A first glimpse at the incompatibility can be seen in the middle panels of Figure \ref{fig.embeddings}. In a previous work \parencite{Gonzalez2019} we demonstrated that, although asynchrony $e_n$ can be precisely defined and measured in operational terms and it is the most fundamental observable in an isochronous paced finger-tapping experiment \parencite{Chen1997}, it is not suitable as a modeling variable in situations involving period perturbations. That's why we plot  the related variable ``predicted asynchrony'' $p_n$ (see \nameref{sec.methods}). This variable is well-defined in the presence of period perturbations and allows us to interpret the evolution of the time series  with the tools of Dynamical Systems theory \parencite{Gonzalez2019}. Predicted asynchrony $p_n$ in Figure \ref{fig.embeddings} shows a divergence between PS (blue) and SC (orange) at $n=1$ in the pure context conditions despite having the same value at $n=0$, whereas in the combined context PS and SC don't diverge from each other.

The divergence of $p_n$ between PS and SC in pure context is a clear hint about incompatibility between them: since the stimulus sequence was the same up to and including $n=0$, and the state of the system is the same at that point (same value of variable) but they show different futures ($n=1$), then they must come from different underlying dynamical systems.

This is further supported by analyzing the whole resynchronization of $p_n$. To show this we reconstructed the experimental phase space from each time series using a difference embedding \parencite{Gilmore1998,Gonzalez2019} (see \nameref{sec.methods}). The results are shown in the right panels of Figure \ref{fig.embeddings}. In the pure context (top right panel), the trajectories in the upper left quadrant associated with SC and PS perturbations intersect each other. This holds true for both small and large perturbations. This is clear evidence of incompatibility between them, since the flow departing from the intersection points in different directions. In a deterministic 2D dynamical system with no hidden variables (like all implementations of the two-process mechanism), a single point in phase space can’t have two different futures \parencite{Bavassi2013}. We emphasize that this result is independent of any particular mathematical model, whether linear or nonlinear.

On the other hand, in the combined context (bottom right panel), the flows exhibit a smooth organization, with trajectories that do not drastically intersect as before. It is worth noting that the data for the 20 ms perturbation exhibit high variability \parencite{Silva2024}, which could explain a slight overlap at noise level for this perturbation size. Considering this, it can be concluded that data in the combined context suggest a common dynamics (for both small and large perturbations). This shows the plausibility of a shared underlying solution in the combined context capable of simultaneously describing the responses to all perturbation types, signs, and sizes.

In summary, these results show that responses after SC and PS perturbations in pure contexts are incompatible with each other. While in the combined context the flow accommodates smoothly, the pure contexts reveal specialized responses that cannot be reconciled under a single underlying mechanism. This observation has methodological implications for future studies of sensorimotor synchronization and provides empirical evidence for an underlying mechanism that can represent diverse types, sizes, and signs of perturbation with a single set of terms and parameters that is nonetheless calibrated by the statistical history of the environment.

\subsection{\label{sec.fit_pure_separate_1}Mathematical models fitted to pure conditions are incompatible with each other}

To show how the experimental incompatibility described in the previous subsection manifests itself in a mathematical model, we consider the following nonlinear dynamical system (see \nameref{sec.methods}):
\begin{equation*}
\begin{aligned}
p_{n+1} & = a e_n + b (x_n - T_n) + \beta e_n (x_n - T_n)^2 \\
x_{n+1} & = c e_n + d (x_n - T_n) + \delta e_n^2 + T_n
\end{aligned}
\end{equation*}

\noindent and fit it to the data. To test the incompatibility hypothesis between perturbation types in pure contexts, two independent fittings were performed, grouping the experimental data according to perturbation type (SC pure vs PS pure; see \nameref{sec.methods}).

If the incompatibility hypothesis were true, we should find that the fittings yield incompatible phase spaces, whereas if it were false, the fitted models should have flows with compatible dynamics.

\begin{figure}[!pt]
\includegraphics[width=\textwidth]{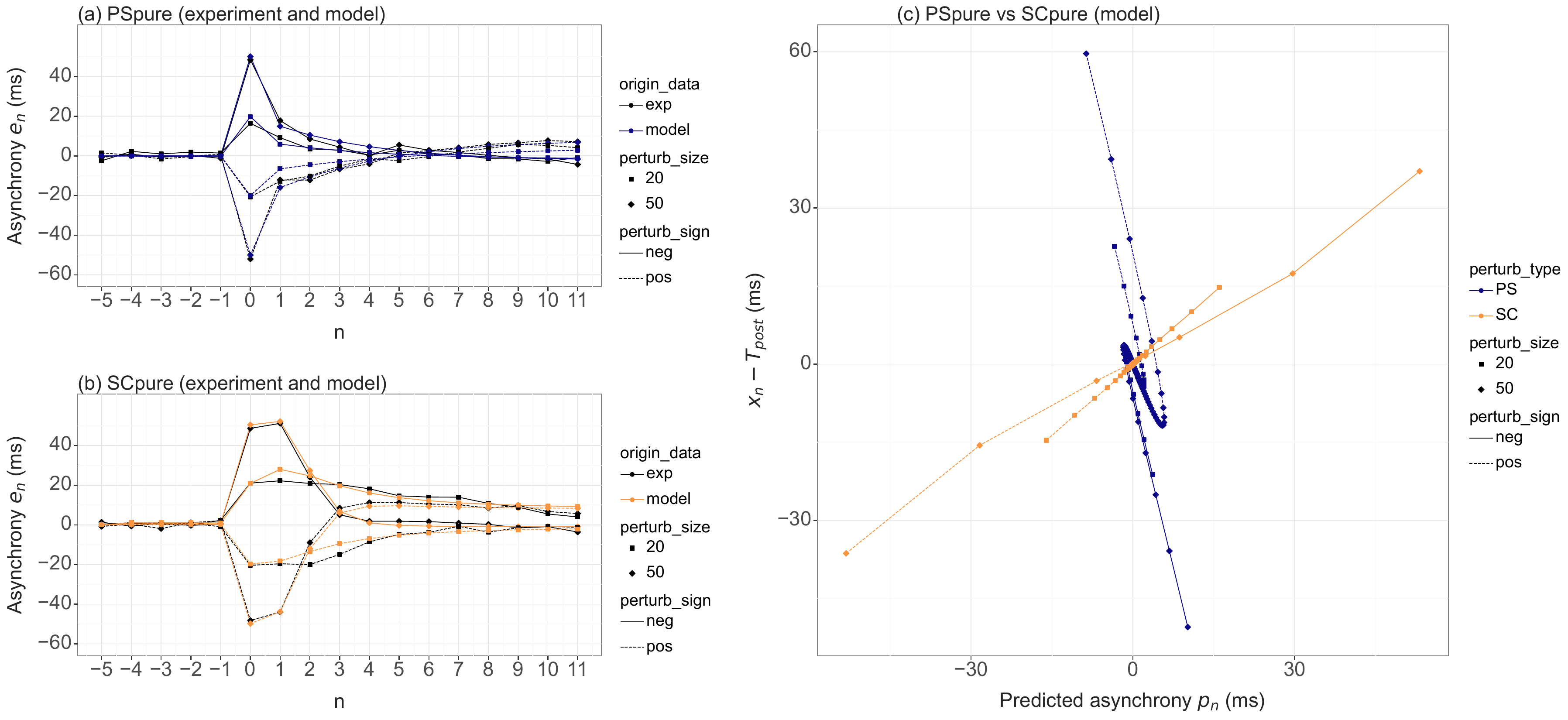}
\caption{Model fitting to the incompatible experimental time series leads to model phase spaces that are also incompatible to each other. (a) and (b): Separate model fittings (color) to SC and PS experimental data from pure contexts (black). (c) Phase space of the fitted models (SC: $1 \leq n \leq 8$; PS: $2 \leq n \leq 8$). Incompatibility is manifested as clear crossings between trajectories of different colors.
\label{fig.fit}}
\end{figure}

Results are plotted in Figure \ref{fig.fit}. The left panels show the experimental time series (black) and the corresponding fitted model time series of PS in pure context (blue) and SC in pure context (orange). Selected trajectories from the models' phase spaces are plotted on the right panel. It is clear that the trajectories of SC and PS perturbations cross each other, pointing to incompatible underlying systems.

\subsection{Parameter distributions of different perturbations in pure context do not overlap}
\label{sec.fit_pure_separate_200}

To quantitatively evaluate the incompatibility hypothesis, we applied the fitting algorithm 200 times to each pure condition independently (SC in pure context and PS in pure context). Our goal was to compare the resulting parameter distributions and determine whether the overlap between them is negligible (further supporting the idea of incompatibility, meaning there are no shared parameter values) or, on the contrary, they clearly overlap (i.e., they are compatible, since they share possible parameter values).

\begin{figure}[!pt]
\includegraphics[width=\textwidth]{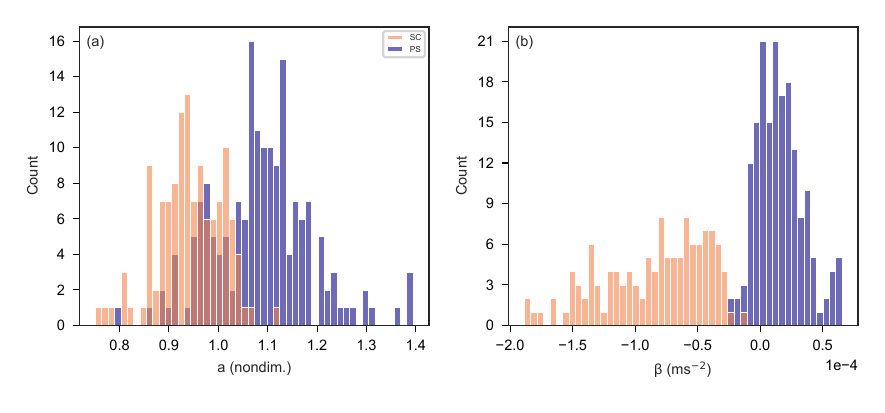}
\caption{Separate model fittings to SC in pure context and PS in pure context don’t share common parameter values for all parameters. (a) There is some overlap between the obtained distributions of parameter $a$. (b) Distributions of parameter $\beta$, however, show negligible overlap.
\label{fig.fit_pure_sep_200}}
\end{figure}

The resulting distributions are shown in Figure \ref{fig.fit_pure_sep_200}. It is evident that, for parameter $a$, there is a clear overlap between SC in pure context and PS in pure context, which might suggest the plausibility of a common solution. However, the overlap is negligible for parameter $\beta$, ruling out the possibility of finding a single set of parameters that could reproduce both types of perturbation unless it is narrowly tuned.

To quantify the extent of overlap between the obtained distributions, we defined the statistic $O_{obs}$ (see \nameref{sec.methods}) that measures how much two distributions empirically overlap based on the range of values they both share and, therefore, that would allow a joint fit of the model. Whereas the distributions of parameter $a$ in Figure \ref{fig.fit_pure_sep_200} have a large overlap ($O_{obs}=71.5\%$), the distributions of parameter $\beta$ display a very small overlap ($O_{obs}=3.7\%$). This allows us to conclude that a narrow model tuning should be necessary to reproduce both SC in pure context and PS in pure context with a single set of parameter values, contrary to the common observation of this very robust synchronization behavior.

\begin{figure}[!pt]
\includegraphics[width=\textwidth]{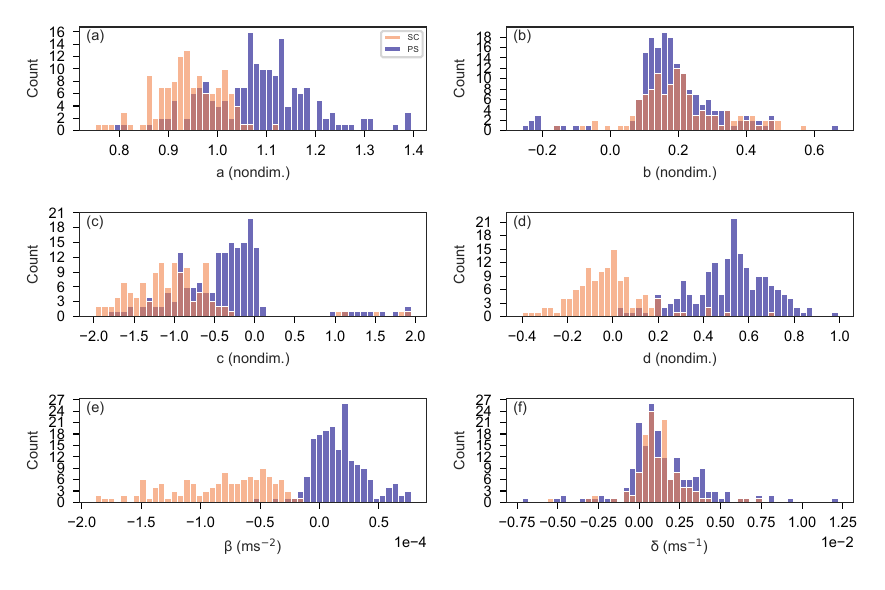}
\caption{All six parameter value distributions from separate model fittings to SC in pure context and PS in pure context. Whereas most parameters show clearly overlapping distributions, overlap for parameter $\beta$ is negligible.
\label{fig.distributions3}}
\end{figure}

This result theoretically supports the hypothesis of incompatibility between SC and PS in pure contexts, previously proposed based on the experimental incompatibility observed in Figure \ref{fig.embeddings}(b). Consequently, we conclude that, in the pure context, the model requires separate fits for each type of perturbation. See the distributions of all parameters in Figure \ref{fig.distributions3}.

\subsection{Parameter distributions from different perturbations in a combined context do overlap}
\label{sec.fit_comb_separate_200}

The counterpart to the incompatibility hypothesis between SC perturbations and PS perturbations in pure contexts is that when these perturbations occur in a combined context instead, they are indeed compatible, as shown experimentally in Figure \ref{fig.embeddings}(d). To further support the incompatibility hypothesis, then, we should show that a similar analysis but with combined context data results in the fitted parameter distributions showing overlap between PS and SC for all model parameters.

\begin{figure}[!pt]
\includegraphics[width=\textwidth]{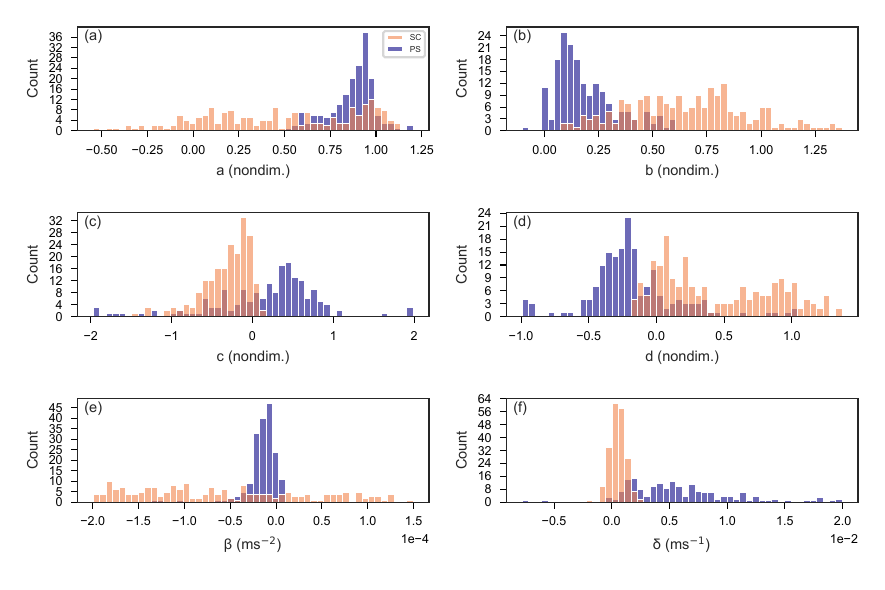}
\caption{Parameter value distributions from SC and PS in combined context. Separate fittings for SC and PS lead to overlapping distributions for all six model parameters.
\label{fig.fit_comb_sep_200}}
\end{figure}

Figure \ref{fig.fit_comb_sep_200} shows the results. We performed separate model fittings to SC and PS from combined contexts this time and their distributions overlap for all parameters. This result reinforces the hypothesis of solution compatibility between SC and PS in combined context, previously proposed based on the flow compatibility observed in Figure \ref{fig.embeddings}(d). This means that we can find a single set of model parameter values that reproduce both types of perturbation at once, with no fine tuning needed. As in the previous section, the extent of overlap was quantified using $O_{obs}$. All parameter distributions show a large overlap between SC and PS, with a minimum overlap of 64\% (panel d) and a maximum overlap of 73\% (panel a).

\subsection{Joint fitting: pure context fares worse than combined context}
\label{sec.fit_joint}

An additional piece of converging evidence regarding the incompatibility between SC and PS in pure contexts comes from the joint (rather than separate) fitting of experimental conditions. By joint fitting we mean fitting a single set of parameters to both perturbation types at once. If SC in pure context and PS in pure context are shown to be incompatible when separate fittings are performed while in the combined context they are compatible (see the two subsections above), then an even stronger case would be built by showing that joint fitting of SC and PS from pure contexts is much worse than joint fitting of SC and PS from combined context.

\begin{figure}[!pt]
\includegraphics[width=\columnwidth]{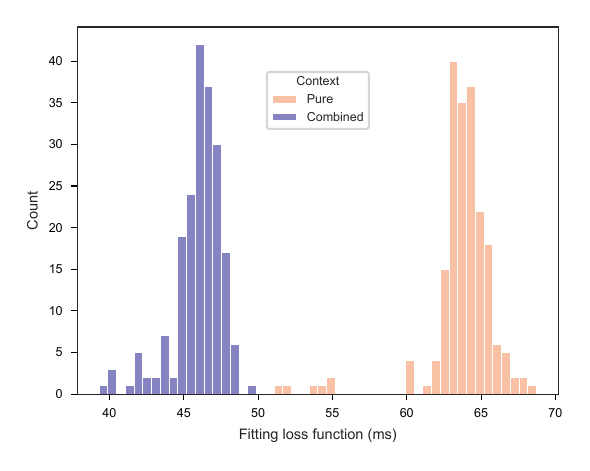}
\caption{Joint fitting of SC and PS perturbations from pure context show lower-quality fits (i.e., larger values of the fitting loss function) than its counterpart from combined context.}
\label{fig.distance}
\end{figure}

We performed 200 independent joint fittings for each condition (that is, 200 joint fittings of SC and PS in pure context, and 200 joint fittings of SC and PS in combined context) and compared the quality of the fitting by looking at the resulting values of the fitting loss function (the objective function to be minimized, chosen to be the Euclidean distance between times series, $\mathit{dist}$; see \nameref{sec.methods}). Figure \ref{fig.distance} shows that the distances obtained in the combined context are systematically smaller than those in the pure context, with two distinct distributions. All distance values in the pure context were greater than 51 ms with a median of 63 ms, while all distance values in the combined context were less than 49 ms with a median of 46 ms. To assess statistical significance, we ran a non-parametric Mann-Whitney-Wilcoxon test resulting in significant differences between pure and combined context distributions with a very large effect size ($U=39402$, $p=10^{-10}$, Cliff's delta $=1$, mean of Cliff's delta by bootstrap $=1$ with 95\% CI $=[1, 1]$).

Since a smaller distance indicates a better model fit (i.e., a smaller value of the fitting loss function), these results mean that the joint fitting  of both types of perturbation in the combined context is closer to the experimental time series than in the pure context. In other words, we interpret this result as the time series in the pure context being more difficult to fit at once because they stem from incompatible systems.

An example of compatible trajectories in phase space for a model from a joint fitting in combined context is shown in Figure S1 in Supplemental Material. The trajectories following PS and SC perturbations join smoothly as they approach the post-perturbation equilibrium, and the predicted asynchrony $p_n$ doesn't show diverging futures.

\subsection{\label{sec.correlations}Parameter correlations and model overparameterization}

The pairwise joint distributions of fitted parameter values (Figures S2 and S3 in Supplemental Material) show that there are some correlations between them, especially among the linear parameters, suggesting that the model is overparameterized. This behavior suggests that the results could be adequately explained by an even simpler version of the model where some of the linear parameters are expressed as a function of others, thus reducing the degrees of freedom. A slight correlation between linear and nonlinear parameters was also observed in some cases.

For the joint fitting of PS and SC in pure context in particular, no correlation between parameters was observed. The absence of correlations implies that the model needs to use all parameters independently to reproduce the data, in contrast to the individual fittings within the same context (SC in pure context and PS in pure context). That is, the model capabilities are strained to their limits and, even in that case, the fitting has a low quality (large fitting loss function values). On the other hand, the nonlinear parameters ($\beta$ and $\delta$) do not show clear correlations with each other, which indicates that the nonlinear part is appropriately parameterized.

A criticism could be raised that, if the model is overparameterized, then any correlation between parameters would prevent us from concluding about incompatibility. This is not true, as we show in the Supplemental Material 

\section{Discussion and Conclusion}

\subsubsection*{Basic findings}

In this work we showed that responses to period perturbations of different types in paced finger tapping (step changes SC vs phase shifts PS) are dynamically incompatible when they occur in different experiments. This led us to conclude, both empirically and theoretically, that they can't be produced by a single underlying error-correction mechanism, and this is due to the existence of a previously shown context effect \parencite{Silva2024}. When they occur in the same experiment, however, they are compatible with each other and a single mathematical model can be successfully fitted to the whole dataset. We interpret these results as mounting evidence about the plausibility of a unique underlying nonlinear mechanism supporting the resynchronization behavior for all perturbation sizes, signs, and types, which nonetheless is tuned to the specific temporal context of an experiment. A summary of the converging evidence supporting our claims can be found in Table \ref{tab.context_summary}.

We reached our conclusions by specifically analyzing the resynchronization phase, that is the transient part of the response between the preperturbation and postperturbation baselines. As we previously showed \parencite{Silva2024,Bavassi2013,Lopez2019,Gonzalez2019}, many defining aspects of behavior are only evident in this information-rich transient. Building on this research programme, the present study further establishes the correct way of comparing the transients when they come from different perturbation types—we need to control for the perturbation context.

\subsubsection*{Conflicting experimental data explained}

Our results offer a unified view of conflicting experimental data in the finger tapping literature: the timing of the first response after a perturbation is different depending on the perturbation type, despite that the stimulus sequence up to and including the perturbation is the same. This fact has been interpreted as due to the participant’s “intentions and expectations” \parencite{Repp2002} after noting different phase correction responses (PCR) to phase-shift perturbations and event-onset shift perturbations. We emphasize that our results show that these different ``expectations'' are not mere quantitative changes to some correction coefficient but are represented and produced by qualitatively incompatible dynamical systems. For more examples of conflicting experimental data and the few existing papers with combined context \parencite{Large2002}, see a literature review on SC and PS perturbations in Ref.\ \parencite{Silva2024}. At the same time, our account offers an explanation for the absence of model fittings in the finger tapping literature representing different perturbation types with a single set of parameter values---if the data are incompatible, then no single model can satisfactorily fit them all. Even our own previous work \parencite{Bavassi2013} showed some modelling difficulties in order to get high quality model reproduction and predictions by using a single model across all perturbation types, signs, and sizes. This is why the usual (and less parsimonious) way in the literature is to fit a model separately for every experimental condition \parencite[see discussion in][]{Gonzalez2019}.

\subsubsection*{A single, nonlinear underlying mechanism}

Our results challenge the proposed idea of ``different mechanisms for different perturbation types'' \parencite{Repp2005,Repp2013}: that is, the proposed existence of two different processes, named phase correction and period correction, that are selectively engaged depending on the perturbation type. According to this proposal, the phase correction process only is active in PS perturbations, while both the phase correction and period correction processes are active in SC perturbations \parencite{Repp2004,Repp2005,Repp2013}. These processes are, in addition, associated with separate model equations. Our results from the combined context, showing that the SC and PS experimental data are compatible with each other and are both well fitted by a single model with both model equations at play, warrant a revision of this idea. When participants are exposed to both perturbation types at random in the same experiment, they don’t seem to select different strategies but a single, intermediate strategy \parencite{Silva2024} represented by a unique set of model parameter values.

Our model can be thought of as a nonlinear implementation of a two-process mechanism. Interpretation of linear coefficients $a,b,c,d$ would follow traditional rules \parencite{Repp2005}, with variable $x$ as the internal representation of the interstimulus interval $T_n$ \parencite{Gonzalez2019,Bavassi2013}. Nonlinear coefficient $\delta$ represents a second-order correction based on the past asynchrony value: the second row in Eq.\ \ref{eq.model_methods}, that is $x_{n+1} = c e_n + \delta e_n^2 + \ldots$, can be rewritten as $x_{n+1} = (c + \delta e_n)e_n + \ldots$ thus making $(c + \delta e_n)$ a correction coefficient that itself depends on asynchrony. Nonlinear term $\beta e_n x_n^2$ might represent an interaction of the two percepts $e_n$ and $x_n$ potentially due to shared resources \parencite{Gonzalez2019,Bavassi2013}.

We propose that during resynchronization after either perturbation type both processes are simultaneously engaged, with a fixed set of parameter values that is chosen among three alternatives depending on the condition: combined context vs phase shift pure context vs step change pure context \parencite{Silva2024}. Based on this, we further emphasize the notion that the putative process being studied in a specific experiment (phase correction or period correction) is not to be identified with the perturbation type used (phase shift or step change)---if the putative phase and period correction processes have distinct neural correlates, an association with a particular perturbation type would rather confound them instead of isolating them.

\subsubsection*{Quantitative versus qualitative results}

Our main result is not quantitative but qualitative in nature: we plot trajectories from different perturbations in pure contexts and show that they cross each other in phase space. It doesn’t matter “how much” they cross—they just cross transversally, instead of smoothly merging as it would be expected if they were produced by any single dynamical system. We note the strength of our qualitative result: there is probably no experimental sample size, however large, and no model fitting, however accurate, that would make the crossing go away. It points to a fundamental difference between conditions beyond the statistically significant difference observed in our previous work \parencite{Silva2024}. The analysis of the geometrical organization of trajectories in phase space is a standard tool in the theory of Dynamical Systems to tell systems apart. Relevant to our work is the fact that a deterministic, two-dimensional model with no hidden variables, like most SMS models of this kind including all implementations of the two-process mechanism, cannot display crossings between different trajectories---all trajectories should merge smoothly (outside of fixed points). A crossing would mean that the values at the crossing could lead to two different futures, thus violating the unicity of solutions \parencite{Bavassi2013,Gilmore1998}. This conceptualization allows us to conclude, within the assumptions of the two-process framework itself, that the pure context conditions are incompatible and thus produced by different dynamical systems regardless of whether they are linear or nonlinear.

\subsubsection*{Potential alternative explanations}

We can picture two other potential explanations alternative to our own. Both would call for a departure from the traditional two-dimensional modeling of the two-process mechanism. The first one is that the trajectory crossings in phase space in Figure \ref{fig.embeddings} could be resolved just by augmenting the dimensionality of the system by adding a third, ``hidden'' variable so that the apparent crossing has actually different values of the third variable. Yet, this would call in addition for an interpretation of the additional variable and even more parameters. In any case, either dynamical incompatibility or the existence of a third additional variable, our results suggest that a reconsideration of traditional theoretical accounts is needed.

Related to this alternative explanation, our experiment was designed as usual in the literature with a common pre-perturbation baseline interstimulus interval $T=500$ ms from which all period perturbations start. The reason was to make the pre-perturbation stimulus sequence equal among conditions in order to specifically test our hypothesis. This unavoidably produced different post-perturbation interstimulus intervals. A theoretical prediction based on our results is that, in a more complex experimental design where all post-perturbation interstimulus intervals are equal we will still find the dynamical incompatibilities as crossings in phase space. The experimental design would need to be more complex (and double duration) because the interesting conditions would be $T=550 \mbox{ ms} \rightarrow 500 \mbox{ ms}$ and $T=450 \mbox{ ms} \rightarrow 500 \mbox{ ms}$ (same post-perturbation ISI) but in addition we would need the sham conditions $T=550 \mbox{ ms} \rightarrow 600 \mbox{ ms}$ and $T=450 \mbox{ ms} \rightarrow 400 \mbox{ ms}$ for the participant not to predict the upcoming perturbation sign.

A second alternative explanation involves construing the underlying correction mechanism as an adaptive, augmented dynamical system whose parameter values in turn evolve over a slower time scale. This alternative explanation is related to Neural Resonance Theory \parencite{Harding2025} and the concept of attunement \parencite{Tichko2022}, where oscillator properties gradually adapt in response to stimulation. While evolution of asynchrony would still be defined on a tap-by-tap basis, evolution of parameters would reflect integration of context information on a trial-by-trial basis. Within the assumptions of the traditional two-process framework, however, this would mean either our conclusions are right or the two-process framework needs at least one additional process (and at most as many as dynamically changing parameters) to account for context.

\subsubsection*{Neural correlates and mathematical representation of sensory expectation}

In a version of the reaching task, it was recently demonstrated \parencite{Michaels2025} that when humans and monkeys are probabilistically cued about the direction of an upcoming mechanical perturbation, they incorporate this information to correct and improve their responses. Importantly, when this information was accumulated over trials, brain areas related to motor preparation and execution showed neural correlates of this information, suggesting that at least the motor aspects of the underlying mechanism change their neural representation depending on context. The accumulation of sensory expectation occurs very similarly in our experimental data, where the perturbation type was implicitly cued as the participants were not told what perturbation type they would be presented with. Although research on the neural correlates of SMS has seen a surge in recent years \parencite[see Discussion in][]{Silva2024}, there are still many unexplained fundamental aspects to it and neural-based mathematical models are lagging behind. Our work shows that sensory expectations can be explicitly and quantitatively described by a single set of parameter values of a behavioral mathematical model that reproduces all data in an experiment.

\subsubsection*{Quantitative tools supporting clinical assessment of motor timing deficits}

Clinical assessment of fine motor function typically involves measures of motor control that, although based on standardized rating scales, are subjective \parencite{Roalf2018}. On the other hand, although some neuropsychological assessment tools are quantitative, they measure only very basic quantifiable aspects of timing behavior. For example, the Finger Tapping Test (FTT, also known as Finger Oscillation Test) in the Halstead-Reitan neuropsychological test battery \parencite{Reitan2003} measures the number of taps a patient performs in a fixed 10-second interval. Traditional measures from the finger tapping literature like average asynchrony and standard deviation, and intertap intervals \parencite{Repp2013} are absent, and research on their diagnostic capabilities is scarce \parencite{Roalf2018}. Models are even less used \parencite{Lainscsek2012}. Our work represents one more step towards understanding the underlying mechanism so appropriate quantitative measures of behavior can be defined to distinguish health from disease.

\begin{table*}[!pht]
  \vspace*{2em}
  \begin{threeparttable}
    \caption{Summary of converging evidence supporting the incompatibility hypothesis between SC and PS perturbations in pure context, and compatibility in combined context.}
    \label{tab.context_summary}
    \begin{tabular}{@{}p{2.8cm}p{4.8cm}p{4.4cm}p{4.4cm}@{}}
      \toprule
      &
      \multirow{2}{*}{\parbox{4.5cm}{\textbf{Experimental embedding\\and Model phase space}}} &
      \multicolumn{2}{c}{\textbf{Parameter distributions}} \\ \cmidrule{3-4}
       & &
       \textbf{Separate fitting} &
       \textbf{Joint fitting} \\
       \midrule

      \textbf{Pure context}
      &
      \makecell[l]{
        \textbf{SC and PS incompatible} \\
        (trajectories cross; \\
        Figs.~\ref{fig.embeddings}b and \ref{fig.fit}c)
      }
      &
      \makecell[l]{
        \textbf{SC and PS incompatible} \\
        (non-overlapping distributions; \\
        Fig.~\ref{fig.fit_pure_sep_200})
      }
      &
      \makecell[l]{
        \textbf{SC and PS incompatible} \\
        (low-quality fitting results; \\
        Fig.~\ref{fig.distance})
      } \\[1em]
        \midrule
      \textbf{Combined context}
      &
      \makecell[l]{
        \textbf{SC and PS compatible} \\
        (smooth trajectories; \\
        Fig.~\ref{fig.embeddings}d)
      }
      &
      \makecell[l]{
        \textbf{SC and PS compatible} \\
        (overlapping distributions; \\
        Fig.~\ref{fig.fit_comb_sep_200})
      }
      &
      \makecell[l]{
        \textbf{SC and PS compatible} \\
        (high-quality fitting results; \\
        Fig.~\ref{fig.distance})
      } \\
      \bottomrule
    \end{tabular}
  \end{threeparttable}
\end{table*}

\subsection*{Constraints on generality}

We expect that our main result, that is the dynamical incompatibility between responses from different temporal contexts created by step changes and phase shifts, will hold when considering event-onset shifts, the third most used perturbation type in the literature \parencite{Repp2005,Bavassi2013}. The extent of the incompatibility, however, cannot be extrapolated from our results and will require experimental determination. Similarly, it cannot be assumed without further evidence that our results will generalize to other sensory modalities (e.g., visual), as stimulus and feedback modalities affect aspects of SMS like the point of subjective synchrony \parencite{Repp2005,Aschersleben2001}. Whether dynamical incompatibility exists for other modalities remains an open empirical question.

Within the two-process mechanism framework, it was shown by using a dual-task paradigm (diverted attention) that the putative period correction process depends on the attentional requirements of the task, but phase correction doesn't \parencite{Repp2004,Repp2005}. On the other hand, we have previously shown that attention can be specifically oriented to the temporal aspects of the task, at least in a pure SC context \parencite{Versaci2021}. This leaves the open question of what the role of attention is in a combined context, where we showed that the two putative processes are engaged at the same time. We predict that attention will have the same effect on all responses within a context, probably with a differential effect depending on which context, but this can't be directly extrapolated from our results and should be experimentally determined.

\section*{CR\lowercase{edi}T author contributions}

A.D.S.: Data curation; Formal analysis; Investigation; Software; Visualization; Writing–original draft; Writing–review \& editing. C.R.G.: Formal analysis; Investigation; Writing–review \& editing. R.L.: Conceptualization; Formal analysis; Funding acquisition; Methodology; Supervision; Visualization; Writing-review \& editing.

\section{Supplemental material}

Supplemental material and figures are available on OSF (\url{https://osf.io/bazhp}) and GitHub (\url{https://github.com/SMDynamicsLab/Incompatibilities2026}).

\printbibliography

\end{document}